\documentclass[sigconf]{acmart}

\begin{document}

\title{Knowledge-enhanced Session-based Recommendation with Temporal Transformer}
\author{Rongzhi Zhang${^1}$,Yulong Gu${^1}$,Xiaoyu Shen${^2}$, Hui Su${^3}$}
\email{zhangrzpanda@gmail.com,  guyulongcs@gmail.com}
\affiliation{%
  \institution{$^1$Alibaba \, Group, China\\
  $^2$Saarland \, Informatics \, Campus, Germany\\
  $^3$Pattern\, Recognition\, Center, Tencent, China}
}


\begin{abstract}
Recent research has achieved impressive progress in the session-based recommendation. 
However, information such as item knowledge and click time interval, which could be potentially utilized to improve the performance, remains largely unexploited.
In this paper, we propose a framework called \textbf{K}nowledge-enhanced \textbf{S}ession-based Recommendation with \textbf{T}emporal \textbf{T}ransformer (KSTT) to incorporate such information when learning the item and session embeddings. 
Specifically, a knowledge graph, which models contexts among items within a session and their corresponding attributes, is proposed to obtain item embeddings through graph representation learning. We introduce time interval embedding to represent the time pattern between the item that needs to be predicted and historical click, and use it to replace the position embedding in the original transformer (called temporal transformer). The item embeddings in a session are passed through the temporal transformer network to get the session embedding, based on which the final recommendation is made.
Extensive experiments demonstrate that our model outperforms state-of-the-art baselines on four benchmark datasets.
\end{abstract}
%



\maketitle

\section{Introduction}
Session-based recommendation is a classic problem in recommendation systems~\cite{sasrec}. It aims to predict the next items users will choose based on their historical behaviors \emph{within the current session}. Recently, many researchers have used the idea of Natural Language Processing(NLP) task(like ~\cite{shen2017estimation,devlin2018bert,shen2018nexus,zhao2019unsupervised,qiu2020easyaug,chang2020dart,chang2021neural}) for reference to provide new ideas for session-based recommendation ~\cite{sun2019bert4rec,yuan2020future,xiao2021uprec}. 
For session-based recommendation, the model architecture usually consists of two key components: 
(1) \textbf{Item Encoder} that projects each item into a low dimensional vector. 
In doing so, \citet{contextrec} concatenated the one-hot embedding of the item and its context representation into recurrent neural networks (RNNs).
\citet{SRGNN} further considered the temporal relationships of items. 
(2) \textbf{Session Encoder} which encapsulates all items within the current session into a compact vector. 
\citet{GRU4Rec} proposed a session embedding model based on Gated Recurrent Units (GRUs).
\citet{NARM} and \citet{liu2018stamp} used attention mechanisms to capture users' local and global preferences. 
Recently, 
\citet{GC-SAN} further improved the recommendation performance by utilizing self-attention networks. 


Despite the impressive success, two limitations exist in current models. (1) Item attributes involve important knowledge, and the context information of the item can be obtained by modeling them, which is beneficial to item recommendations.
However, existing approaches ignore them when building the item encoder.  
(2) The session embedding ignores the time interval between behaviors. The time interval between two adjacent clicks in the user's click history is varied. When the time interval is different, the product that the user clicks next should also have a different relationship with the previous behavior. For example, if the user is currently buying beer, the click within 3 minutes may be a beer of a different brand, but after 10 minutes, they may be browsing products such as beer glasses.
Existing self-attention networks like Transformers~\citet{transformer} are powerful at modelling sequential data with position embedding, but they fail to encode the time interval information for predictions, resulting in unsatisfactory results in session-based recommendations.

In this paper, we propose a Knowledge-enhanced Session-based Recommendation with Temporal Transformer (KSTT) to improve the performance of session-based recommendations. In comparison with state-of-the-art methods, our KSTT is equipped with two mechanisms to address the above-mentioned limitations: (1) We build a knowledge graph based on the attributes of the items and the order of clicks between the items, and use the graph neural network (GNN) to obtain the global context information of each node. 
(2) We use a well-designed mapping function to embed the timestamp into k-dimension vectors for modeling the time and time interval information. 
The mapping function is based on recent researches \citep{time2vec,xu2019self}, which replaces position embedding with time embedding to reveal more time patterns and recurrent attention.
In summary, our contributions are listed as follows.
\begin{itemize}
    \item We introduce the knowledge graph into session-based item recommendation, and use GNN to integrate the information of adjacent nodes in the graph.
    \item We introduce temporal transformer into session-based recommendation and testify its effectiveness.
    \item We conduct extensive experiments on four benchmarks datasets to demonstrate that KSTT can significantly overcome state-of-the-art baselines. 
\end{itemize}

\section{KSTT Model}
\begin{figure}[h]
    \centering
    \includegraphics[scale=0.34]{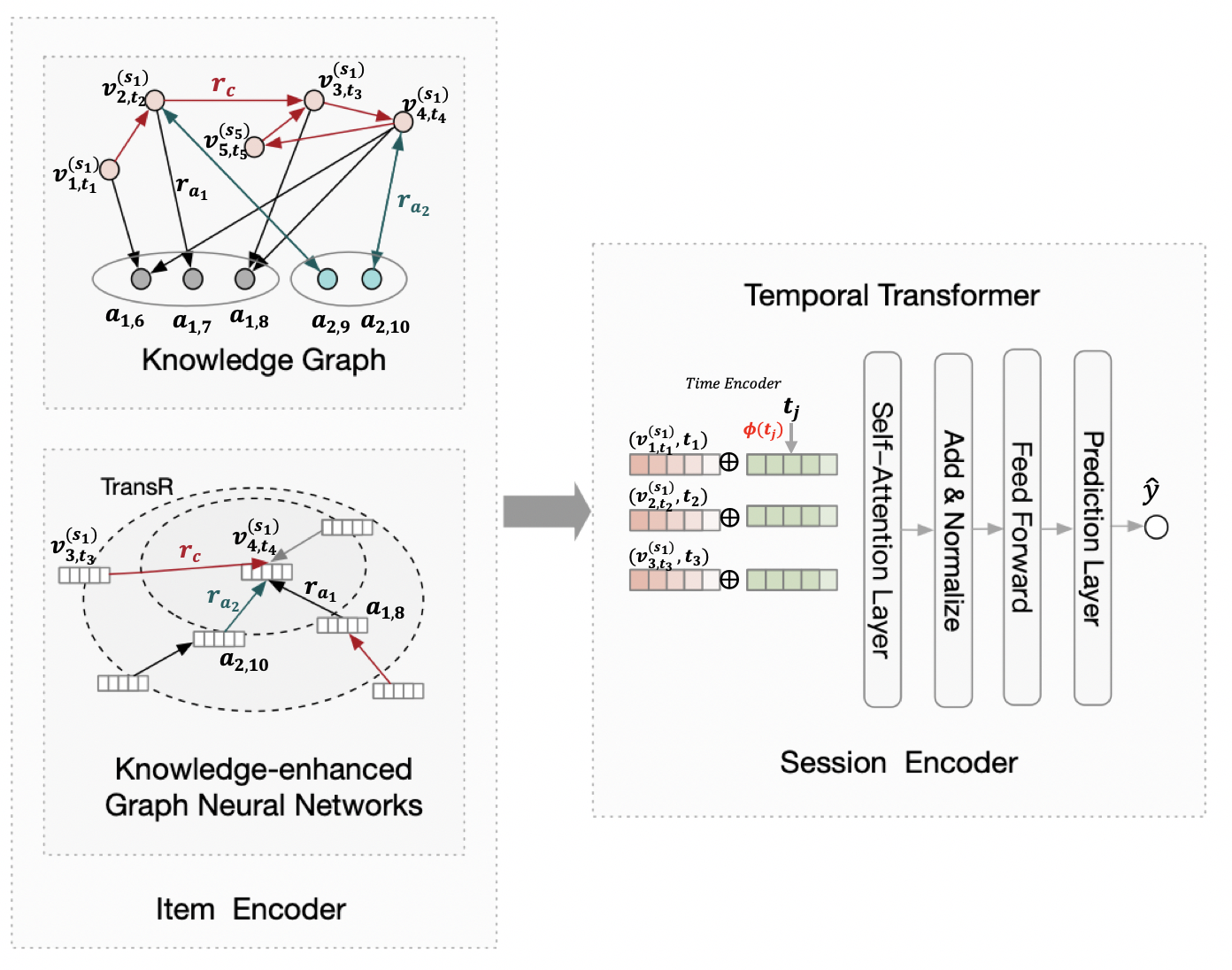} 
    \caption{Architecture of KSTT.}
    \label{fig:structure}
\end{figure}

In this paper, we propose a framework called Knowledge-enhanced Session-based Recommendation  with Temporal Transformer (KSTT) for Session-based Recommendation.  
Given the training set, it has items $\mathcal{V} = \{v_1, v_2, ..., v_M\}$ and attributes $\mathcal{A} = \{a_1, a_2, ..., a_K\}$. Session $s_i=\{v_{i,1},v_{i,2},...,v_{i,t}\}$ means that the $i$-th session contains a transaction with item $v_{i,t}$ at time $t$. 
The architecture of KSTT is demonstrated in Figure ~\ref{fig:structure}. Firstly, the Item encoder in KSTT learns the embeddings of items using Knowledge-enhanced Graph Neural Network. The nodes in the knowledge graph contains two types: items($v$) and items' attributes($a$). The edges in the graph represent the relations($r$) between nodes.
Secondly, the Session encoder in KSTT exploits Temporal Transformer using time encoder $\phi$ to learn session representation. Finally, KSTT calculates the prediction scores $\hat y$ for the next item.
\subsection{Item Encoder with Knowledge-enhanced Graph Neural Network}
The item encoder uses Knowledge-enhanced Graph Neural Network to learn item embeddings based on the knowledge graph.\\
\textbf{Knowledge Graph Construction}
Previous methods~\citep{SRGNN,GC-SAN,su2020moviechats} usually construct the session graph based on the sequential relationship between items. 
In this work, we propose the idea of constructing a directed \textit{KG} based on the relationships in the session as well as the knowledge (e.g. attributes) of items. We construct the \textit{KG}, where the nodes are items and these items' related attributes and the edges are the relationships between the nodes including sequential edges and semantic edges. The sequential edges are denoted as $(v_{i,t},r_c,v_{i,t+1}), 1 \leq t < m$ and c is a fixed number. $r_c$ is the sequential relationship of users' behaviors, and each edge represents that the user clicks the item $v_{i,t+1}$ after consuming item $v_{i,t}$. The semantic edges are denoted as $(v_{i,t},r_j,a_{kj}), 0 < j \leq k+1 $ and 0 < k < K which represent item $v_{i,t}$ and attribute $a_{kj}$ are connected with relation $r_j$.\\
\textbf{Knowledge-enhanced Graph Neural Network}
We exploit our proposed Knowledge-enhanced Graph Neural Network to learn the embeddings of items in the \textit{KG}. We perform the knowledge graph embedding method TransR~\citep{transR} to learn the semantic information and exploit Graph Convolutional Neural Networks~\citep{GCN} to aggregate information from neighbors to capture context information between nodes. 
We use the $\textit{(head,relation,tail)}$ entity triplets to learn the semantic representation of nodes. For each triplet, the embeddings of the head entity $h$, relation $r$ and tail entity $t$ are $d$\-dimension vector updated by graph convolutional neural network~\citep{GCN}. The update function of nodes in one graph convolutional layer can be represented as: $e_{h/t}' = \textit{norm}(\textit{LeakyReLU}(GEW))$, where $e_{h/t}$ is the representation of entity $h$ or $t$. $E$ is the embeddings for all entities and $G$ is the adjacent matrix of knowledge graph. Graph convolution is computed through gathering the response from each neighbour, and multiple graph convolutional layers are stacked to extract high-level representation for nodes. Therefore, graph convolution is suitable for exchanging information in neighborhoods. 
\subsection{Session Encoder with Temporal Transformer}
The Session Encoder exploits Temporal Transformer to capture the relationships among items in a session, and learn the representation of the session.\\
\textbf{Temporal Transformer}
In the Transformer architectures ~\citep{transformer}, position embeddings are used to model the position information in one sequence.
However, for session-based recommendation, this method only models the sequential order instead of time interval information between behaviors. 
In this work, we proposed Temporal Transformer to deal with this issue.
Firstly, we encode each behavior in the session.
For each historical behavior, we calculate the corresponding time interval, which is difference of timestamp between this behavior and prediction time.
Then, we exploit the time encoder to encode the time interval into low dimensional vector.
The embedding of the behavior $v_i$ is defined as $E(v_i)$ , computed by the sum of item embedding $IE_d(v_i)$ and time embedding $TE_d(v_i)$. 
Secondly, as Transformer did~\citep{transformer}, we exploit Multi-head Self-Attention Networks and Feed Forward neural networks(FFN) to learn the representation of the session.
The calculation method is defined as 
$L^{(l)}= \operatorname{FNN}(\operatorname{MultiHead}(L^{(l-1)},L^{(l-1)},L^{(l-1)}))$, 
where $MultiHead(\cdot,\cdot,\cdot)$ is a multi-head self-attentive network that takes a query matrix, a key matrix, and a value matrix as input. $L^{(0)}$ is initialized session representation including behavior embeddings $E(v_i)$ of this session.\\ 
\textbf{Time Encoder}\label{Lab_time_encoder}
We investigate three time embedding methods as time encoder: Time Bucket embedding method, Time2vec~\citep{time2vec} and Mercer time embedding method~\citep{xu2019self}.

\textbf{Time Bucket Embedding method(TBE).}
This method uses log and floor function to discretize time into one-hot representation.

\begin{equation}
  \Phi_{bucket}(t)= \operatorname{One-hot}(\lfloor \operatorname{log_2}(\hat{t} - t) \rfloor)
  \label{eq:timeembedding}
\end{equation}

\textbf{Time2vec(T2v).}
T2v is designed to capture both periodic and non-periodic patterns in time~\citep{time2vec}.
The embedding is calculated using following equation
\begin{equation}
\begin{aligned}
\Phi_{T2V}^i(t) = 
\left\{  
     \begin{array}{lr}
     w_it+b_i,\,         &if\,i = 1\\
     sin(w_it+b_i),\,     &if\,1\leq i < d \\
     \end{array}  
\right.
\end{aligned}
\end{equation}
where $i$ is the dimension.

\textbf{Mercer Time Embedding(MTE).}
Mercer Time Embedding(MTE) learns time representation with transition-invariant property~\cite{xu2019self}.
The detailed Mercer time embedding function is given as follows:
\begin{equation}
\begin{aligned}
\Phi_{MTE}(t) &= [\Phi_{w_1,d}(t),\Phi_{w_2,d}(t),...,\Phi_{w_k,d}(t)]^T\\
\Phi_{w,d}(t) = &[\sqrt{c_1}, \sqrt{c_{2j}}cos(\frac{j\pi t}{\omega}),\sqrt{c_{2j+1}}sin(\frac{j\pi t}{\omega}),...]
\end{aligned}
\end{equation}
where $c$ and $\omega$ represent the parameters. 
$\omega$ should be initialized by uniformly distribution and $\omega_0$ should be initialized as 0. 

In summary, these time embedding methods capture different time patterns within one session which helps attention mechanism to learn session embedding. 
\subsection{Prediction and Optimization}

We adopt the two phases with different loss functions to optimize KSTT.
In one phase, we mainly optimize the loss function based on ~\citet{transR}. After obtaining entity embeddings from knowledge graph, we also utilize a projection matrix $\bold{M}_r$ to project the head entities $\textit{h}$ and tail entities $\textit{t}$ from the entity space to the relation space, represented as $\bold{M}_r\bold{h}$ and $\bold{M}_r\bold{t}$ respectively. Then we try to minimize the difference between $\bold{M}_r\bold{h} + \bold{r}$ and $\bold{M}_{r}\bold{t}$, formulated as the score function $g_r(h,t) = \|\bold{M}_{r}\bold{h} + \bold{r} - \bold{M}_{r}\bold{t}\|^2_2$. A lower score means that the triplet is more likely to be true. We sample head entities from all training nodes with difference relation and obtain a batch containing positive and negative pairs. We minimize the scores of negative pairs and maximize that of positive pairs for learning knowledge graph embeddings as follows.
\begin{equation}
\label{eq:KGLoss}
\ \mathcal{L}_{KG}=\sum_{(h,r,t,{t}')\in T_{KG}}\ln\sigma \left[g_r(h,{t})-g_r(h,t')\right]
\end{equation}
where $T_{KG} =\{(h,r,t,{t}')\mid (h,r,t)\in \mathcal{G}, (h,r,{t}')\notin \mathcal{G}\}$, $(h,r,{t}')$ is a random sampled negative triplet by replacing $t$ with another entity $t'$ in the knowledge base, and $\sigma (\cdot)$ is the sigmoid function.

In the other phase, we aim to predict the next items users will buy. First, we obtain the session representation $\text{s}$ for each user from temporal transformer.
Then we can compute the score $\hat{\mathbf{z}_i}$ for each item $v_i$ by
$	\hat{z} = s_\text{h}^\top \, v.$
we apply a softmax function over $\hat{z}$ and obtain the final output vector $\hat{\mathbf{y}}$ which is the probability distribution of the items. 
we optimize KSTT with the cross-entropy of the prediction $\hat{\mathbf{y}}$ and the ground truth $\mathbf{y}$.
\begin{equation}
\mathcal{L}_{rec} = -\sum_{i = 1}^{n} \mathbf{y}_i \log{(\hat{\mathbf{y}_i})} + (1 - \mathbf{y}_i) \log{(1 - \hat{\mathbf{y}_i})}
\end{equation}
where $n$ is the number of training data. 

We jointly optimize the session-based recommendation and knowledge graph embedding by minimizing the following loss function:
$\label{Sum Loss}
\mathcal{L}_{KSTT}=\mathcal{L}_{rec}+\mathcal{L}_{KG}+\lambda ||\Theta ||^{2}_{2}$,
where $\Theta$ is the parameters of our model, and $\lambda$ is the coefficients of the $L_2$ regularization.

\section{Experiments}
\subsection{Experiment setting}
\textbf{Datasets}
We choose three representative benchmark Yoochoose, Diginetica and Last-fm, to evaluate our model.
 \textbf{Yoochoose} is used as a challenge dataset for RecSys Challenge 2015. 
 It records the click-streams from an E-commerce website within 6 months. As previous did, we split the session sequences by fixed cutoff to generate the train and test datasets, and then further split the train data by the most recent partition 1/4 and 1/64.
\textbf{Diginetica} is used as a challenge dataset for CIKM Cup 2016. 
It contains the transaction data with the basic information of the products and transaction.
\textbf{Last-fm} is a music dataset collected from the Last.fm API, 
which has been widely applied for music artist recommendations. After preprocessing, we keep the top 40,000 most popular artists and filter out sessions that are longer than 50 or shorter than 2 items. \\

\textbf{Baselines}
\label{baselines}
In order to valid the effectiveness of our proposed KSTT model, we compare KSTT with the following representative methods:
\textbf{POP} always recommends the most popular items in the whole training set. \textbf{S-POP} always recommends the most popular items in the current session. \textbf{Item-KNN} computes the similarity between each pair of items with their cosine distance in sessions. 
\textbf{BPR-MF}~\citep{BPRMF} calculates a pairwise ranking loss with a BPR objective function. 
\textbf{FPMC}~\citep{FPMC} is a hybrid model for the next-basket recommendation based on the personalized transition matrix tube. \textbf{GRU4REC}~\citep{GRU4Rec} stacks multiple GRU layers to encode the session sequence into a latent vector, where a ranking loss is used to train the model.
\textbf{NARM}~\citep{NARM} incorporates an attentive network to combine  states of RNN. 
\textbf{STAMP}~\citep{liu2018stamp} uses attention layers to replace all RNN encoders in previous work. 
\textbf{RepeatNet}~\citep{Repeatnet} proposes an encoder-decoder architecture to solve the reputation consumption problem. \textbf{SR-GNN}~\citep{SRGNN} applies a gated graph convolutional layer~\citep{GatedGSNN} to obtain item embeddings. 
\textbf{GC-SAN}~\citep{GC-SAN} introduces a self-attention network to the session encoder.\\ 
\textbf{Parameters and Metric Measure}
The item embedding size and the dropout ratio are set to 100 and 0.1 in all the methods. 
We use Adam as our optimizing algorithm and set the learning rate as $0.001$.
The batch size is set to 246.
To speed up the converge in the training process, we set the mini-batch size as 256 in session graph and 512 in the knowledge graph.\\
For the evaluation metrics, we use R@K (Recall@K) and MRR@K (Mean Reciprocal Rank@K) to measure the recommendation performance. 
\begin{table}[!t]
    \centering
    \caption{Performance comparisons on benchmark datasets.}
    \setlength{\tabcolsep}{0.04cm}
    \scalebox{0.7}{
    \begin{tabular}{cccccccccc}
         \toprule
         &{Method}&\multicolumn{2}{c}{\textit{Diginetica}}&\multicolumn{2}{c}{\textit{Yoochoose1/64}}&\multicolumn{2}{c}{\textit{Yoochoose1/4}}&\multicolumn{2}{c}{\textit{Last-fm}}\\
         &&R@20&MRR@20&R@20&MRR@20&R@20&MRR@20&R@20&MRR@20\\
         \midrule
         &POP&0.89&0.20&6.71&1.65&1.33&0.30&5.26&1.26\\
         &S-POP&21.06&13.68&30.44&18.35&27.08&17.75&22.59&8.71\\
         Classical&Item-KNN&35.75&11.57&51.60&21.81&52.31&21.70&14.84&4.85\\
         &BPR-MF&5.24&1.98&31.31&12.08&3.40&1.57&14.05&5.01\\
         &FPMC&26.53&6.95&45.62&15.01&51.86&17.50&17.68&4.99\\
         \midrule
         &GRU4REC&29.45&8.33&60.64&22.89&59.53&22.60&17.90&5.39\\
         DNN-based&NARM&49.70&16.17&68.32&28.63&69.73&29.23&29.94&10.85\\
         &STAMP&45.64&14.32&68.74&29.67&70.44&30.00&29.24&11.33\\ 
         &RepeatNet& 47.79&17.66&69.13&30.24&70.71&31.03 &32.28&12.03\\
         \midrule
         &SR-GNN&50.73&17.59&70.57&30.94&71.36&31.89&30.75&12.53\\
         GNN-based&GC-SAN&52.48&18.20 &71.20&30.48&71.46&31.47&32.46&12.22   \\
         &KSTT&\textbf{53.25}&\textbf{18.29}&\textbf{72.25}&\textbf{31.05}&\textbf{72.35}&\textbf{32.30}& \textbf{32.63} & \textbf{12.90}\\
         \bottomrule
    \end{tabular}}
    \label{tab:all-baseline}
\end{table}
\subsection{Performance Comparison}
Table~\ref{tab:all-baseline} lists the performance of the proposed model and other state-of-art session-based recommendation methods. 
All of the methods fall into three categories: classical methods, DNN-based methods and GNN-based methods.
Considering POP and S-POP, they ignore the complex relationship between items and make recommendation based on occurrence number. 
Even so, S-POP still outperforms other methods demonstrating the importance of session contextual information. Item-KNN achieves better results than FPMC. Item-KNN utilizes the similarity between items without considering sequential information but MC-based methods often consider adjacent information. These demonstrate that inter-session information is more important than information of neighbors.

In particular, DNN-based methods, explicitly model the users' global behavioral preferences by local and global representation leading to superior performance against traditional methods. GNN-based methods are state-of-the-arts. They model the transactions within a session as a graph, which outperforms other methods significantly. 
Our proposed method KSTT is more powerful to model behaviors in the session and achieves the best performance on all four datasets in terms of R@20 and MRR@20, especially for \textit{Yoochoose} dataset which most have 1.05\% improvement for R@20 and 0.83\% for MRR@20 .

\subsection{Ablation Study}
In this subsection, we conduct the ablation studies to verify the effectiveness of Knowledge-enhanced Graph Neural Network (KGNN) and Temporal Transformer (TT) combining Self-Attention Neural Networks (SAN) and Time Encoder (TE). We conduct the experiments on \textit{Yoochoose1/64} and \textit{Diginetica} and choose SRGNN as baseline. The experimental results are listed in Table~\ref{tab:ablation study}, where $+$ means \textbf{adding} this module or \textbf{replacing} the original similar module on the baseline. We can draw the following conclusions: (1) \textbf{\textit{For KGNN}}: KGNN module indeed helps GNN achieve better performance, evidencing that +KGNN overcomes SRGNN that only uses a session graph. In particular, the KGNN module brings 1.44 (50.73 to 52.17) improvement in R@20 and 0.39 (17.59 to 17.88) gains in MRR@20 on \textit{Diginetica} and the improvement on \textit{Yoochoose1/64} is 0.26 and 0.17, respectively. Since \textit{Diginetica} has more attributes than \textit{Yoochoose}, the former has more knowledge than the latter. Therefore, we can conclude that more knowledge information can bring better improvement.
(2) \textbf{\textit{For TT}}: We use three time embedding functions. The results show that all the time representation methods help session encoder with more information, especially TBE. This demonstrates that TBE has superiority the inter-session time pattern for session-based recommendation.

\begin{table}[!t]
    \centering
    \caption{Ablation study of modules.}
    \scalebox{0.8}{
    \begin{tabular}{cccccc}
         \toprule
         {Models}&\multicolumn{2}{c}{\textit{Yoochoose1/64}}&\multicolumn{2}{c}{\textit{Diginetica}}\\
         &R@20&MRR@20&R@20&MRR@20\\
         \midrule
		 Baseline & 70.48 & 30.81& 50.73 & 17.59 \\
		 Baseline (+KGNN) &70.74 & 30.98 & 52.17 & 17.88 \\ 
		 Baseline (+SAN) & 71.20 & 30.48 & 52.48 & 18.20 \\ 
		 Baseline (+KGNN, +SAN) & 71.53 & 30.50 & 53.10 & 18.25 \\
         \midrule
         Baseline (+KGNN, +SAN, +MTE) & 72.09 & 30.63 & 53.12 & 18.24 \\
         Baseline (+KGNN, +SAN, +T2v) & 72.10 & 30.70 & 53.15 & 18.27 \\
         Baseline (+KGNN, +SAN, +TBE) & \textbf{72.25} & \textbf{31.05} & \textbf{53.25} & \textbf{18.29} \\
         \bottomrule
    \end{tabular}}
    \label{tab:ablation study}
\end{table}
In summary, our proposed modules can significantly improve the baseline model, which result in the state-of-the-art performance.

\section{Related Work}
\subsection{Session-based Recommendation}
Session-based recommendation, which is an emerging topic in the recommendation system, has attracted interests of many researchers from both academia and industry.
\citet{GRU4Rec} use a gated recurrent unit model to model the user's behaviors in each single session. 
\citet{NARM} utilizes attention mechanisms to fusion the local interests and global users' preference.
\citet{liu2018stamp} further takes the long/short term memory into account when constructing a neural attention model.

\subsection{Graph Neural Network for Session-based Recommendation}
%
Graph Neural Networks(GNN), which can capture both graph structure and nodes' attributes in the graph, 
have shown its superiority in many applications.
\citet{session-basedsocialrec} represents the interactions between users and items as a bipartite graph.
\citet{GatedGSNN} proposed Gated-GSNN for sequential learning in graph structure.
\citet{SRGNN}, which constructs item-item graph in each session, is a pioneering work that first use GNN to capture the nonlinear relationship between items in a session.
It generates session embedding using attentive networks and predict the next item. 
~\citet{GC-SAN} uses transformer to learn inter-item dependencies and generate session embedding for prediction.

\section{Conclusion}
In this paper, we present a novel architecture that incorporates knowledge graph and time interval information into session-based recommendation. The proposed method leverages knowledge-enhan\-ced graph neural networks to obtain informative item embedding and uses temporal transformer to learn session representation effectively. Comprehensive experiments show that the proposed approach can consistently outperform state-of-art methods.

\bibliographystyle{ACM-Reference-Format}
\bibliography{ijcai20.bib} 
\end{document}